\documentclass[a4paper,10pt,aps,pre,twocolumn,amsmath,amssymb,longbibliography]{revtex4-1}

\usepackage{graphicx}
\usepackage[usenames]{color}
\usepackage[utf8]{inputenc}
\usepackage{textcomp}
\usepackage{comment}
\usepackage[colorlinks=true,citecolor=blue]{hyperref}
\usepackage{amsmath, amssymb, amsthm, amsfonts} % Almost always needed
\usepackage[arrowdel]{physics} % So you can type stuff like \grad, \div, \laplacian

\usepackage[nameinlink,capitalize]{cleveref}

\newcommand{\ahum}[1]{``#1''}
\newcommand{\olcite}[1]{Ref.~\cite{#1}}

\newcommand{\avg}[1]{\left\langle #1 \right\rangle}

\newcommand{\Nstar}{N^\star}
\newcommand{\PMD}{P_{\rm dis}}

\newcommand{\I}{{\mathrm{i}}}

\begin{document}

\title{Dissipation in solids under oscillatory shear: 
Role of damping scheme \\ and sample thickness}

\author{Richard Vink}
\affiliation{Institute of Materials Physics, Georg-August-Universität Göttingen,
37073 Göttingen, Germany}

\date{\today}

\begin{abstract} We study dissipation as a function of sample thickness in 
solids under global oscillatory shear applied to the top layer of the sample. 
Two types of damping mechanism are considered: Langevin and Dissipative Particle 
Dynamics~(DPD). In the regime of low driving frequency, and under {\it 
strain-controlled} conditions, we observe that for Langevin damping, dissipation 
{\it increases} with sample thickness, while for DPD damping, it {\it 
decreases}. Under {\it force-controlled} conditions, dissipation increases with 
sample thickness for both damping schemes. These results can be physically 
understood by treating the solid as a one-dimensional harmonic chain in the {\it 
quasi-static} limit, for which explicit equations (scaling relations) describing 
dissipation as a function of chain length (sample thickness) are provided. The 
consequences of these results, in particular regarding the choice of damping 
scheme in computer simulations, are discussed. \end{abstract}

\maketitle

\section{Introduction}

Investigations of dissipative processes by means of (classical) Molecular 
Dynamics simulations can provide valuable insights at atomic scale resolution. 
However, the results can be quite ambiguous, since they may depend on the size 
of the sample that was simulated, as well as on the details of the damping 
scheme that was used. Langevin damping is presumably the most commonly used such 
scheme, whereby each atom experiences a friction force whose magnitude is 
proportional to its velocity. One immediate practical problem is choosing the 
Langevin damping parameter, which can significantly affect 
dissipation~\cite{10.1103/physrevb.100.094305, 10.1103/physrevb.82.081401, 
10.1088/0953-8984/22/7/074205}.

However, we emphasize here that Langevin damping is by no means the only choice 
conceivable, and, depending on the system of interest, might not even be 
optimal. One issue with Langevin damping is its violation of momentum 
conservation, as well as a spurious dissipation under global translations of the 
entire system. In this respect, an interesting alternative is the damping scheme 
of {\it dissipative particle dynamics} (DPD)~\cite{10.1209/0295-5075/30/4/001}, 
which does not suffer from these shortcomings (except for the problem of having 
to choose a numerical value of the DPD damping parameter, which still remains). 
While DPD was originally designed to describe complex fluids, it is nowadays 
also being used to describe the electron-phonon coupling in 
metals~\cite{10.1103/physrevlett.120.185501}.

In addition to the damping scheme come finite-size effects, which also affect 
dissipation. In principle, finite-size effects can manifest themselves in 
experimental samples also, and, as such, are not necessarily artifacts. This 
obviously requires that the experimental sample be small in at least one 
dimension, a natural candidate being its thickness. Experiments have indeed 
established that the friction force, under certain conditions, depends on the 
thickness of the sample, the so-called {\it thickness effect}. For layered 
materials, such as graphene, the usual behavior is that friction decreases with 
increasing sample thickness~\cite{10.1038/s41467-019-14239-2}. In 3D~crystals, 
where the lattice planes are strongly bound, the trend appears to be reversed, 
i.e.~friction increases with the sample 
thickness~\cite{10.1209/0295-5075/87/66002, 10.1103/physrevb.82.081401, 
10.48550/arxiv.2210.09677, 10.1209/0295-5075/acd140}. This assumes that the 
lower part of the sample is rigidly fixed: For a free-standing substrate 
(membrane) also the reverse behavior is 
possible~\cite{10.1209/0295-5075/acd140}, but this scenario is not considered 
here.

It is the purpose of this paper to demonstrate, for the case of a 3D~crystal, 
how sensitive the dependence of dissipation on sample thickness in computer 
simulations really is: Depending on damping scheme and driving details, both 
increasing and decreasing behaviors are possible. This shows that great care 
must be taken when comparing thickness effects in simulations to real 
experiments. Essentially, the assumptions of the simulation, including the 
damping scheme, must be argued to resemble experimental conditions. These 
difficulties already arise in relatively simply situations, for example {\it 
low-frequency} and {\it small-amplitude} oscillatory shear, applied to the top 
layer of the crystal. Already here, the dissipated energy can be increasing or 
decreasing with the sample thickness, depending on details.

The physical origin of these different behaviors can be quite easily understood 
from the simple picture of a driven one-dimensional (1D) bead-spring chain, with 
which we begin our paper. We consider Langevin and DPD damping schemes, and 
distinguish between strain- and force-controlled driving scenarios. Next, we 
verify the 1D findings for a 3D system. In addition, some guidelines are 
provided as to how one could choose the numerical value of the damping 
parameters.

\section{1D chain: Scaling laws}

To understand how dissipation under oscillatory shear depends on the thickness 
of the sample, it is instructive to consider a 1D \ahum{bead-spring} chain. In 
this section, we derive \ahum{scaling laws} for the dissipation as a function of 
chain length, in the limit of low driving frequency (or, equivalently, short 
chain length).

\subsection{Undamped chain}

Assume a chain consisting of $i=0,\ldots,N-1$ beads (i.e.~$N$ beads in total) 
positioned on a straight line, the spacing between the beads being~$a$. Each 
bead (mass~$m$) is connected to its left and right neighbors by springs (spring 
constant~$K$). The equation of motion for each bead, retaining only terms linear 
in the bead displacements, then becomes:
\begin{equation}
\label{eq:N2}
m \ddot{u}_i = K(u_{i-1} + u_{i+1} - 2u_i) \,,
\end{equation}
where $u_i \equiv u_i(t)$ is the transversal displacement (SI-unit $[u_i] = \rm m$) of bead 
$i$ at time $t$. Substituting the plane-wave Ansatz, $u_i \propto e^{\I (k i 
a - \omega t)}$, into \cref{eq:N2}, with wavenumber $k=2\pi/\lambda$, wavelength 
$\lambda$, and frequency $\omega$, one easily derives the dispersion relation:
\begin{equation}
\omega = 2 \Omega_0 \sin(ak/2) \, , \quad \Omega_0 \equiv \sqrt{K/m} \,.
\end{equation}
In the long wavelength limit ($k \to 0$), the above dispersion implies a speed 
of sound: $c_0 = \lim_{k \to 0} \omega/k = a \Omega_0$. For metals, $c_0 \sim 
10^3 \, \rm m/s$ and $a \sim \rm \AA$ implying $\Omega_0 \sim \rm THz$.

\subsection{Chain with Langevin damping}

Consider now a damped chain, with the damping (friction) being proportional to 
the bead velocity, which is the standard choice in Langevin dynamics. The 
equation of motion then becomes:
\begin{equation}
\label{eq:N2L}
m \ddot{u}_i = K(u_{i-1} + u_{i+1} - 2u_i) - m \gamma_L \dot{u}_i \,,
\end{equation}
with damping parameter $\gamma_L$, whose SI-unit $[\gamma_L] = \rm 1/s$, 
i.e.~that of frequency. Substituting as before the plane-wave Ansatz, one can 
solve for the inverted dispersion relation $k(\omega)$, i.e.~wavenumber as a 
function of frequency. While an analytical expression can be obtained, it is 
more informative to consider the low-frequency regime:
\begin{equation}
\label{eq:langdamp}
k_L \overset{\omega \ll \Omega_0}{\approx} \begin{cases}
\frac{\omega}{c_0} + \frac{\I \gamma_L}{2c_0} & \gamma_L \ll \omega \,,\\
\frac{ \sqrt{ \gamma_L \omega} }{ c_0 }
\left( \frac{1+\I}{\sqrt{2}} \right) & \gamma_L \gg \omega \,, \\
\end{cases}
\end{equation}
where $\I^2=-1$. The wavenumber is complex, meaning the wave is exponentially 
damped, and only propagates a finite distance $l_L \sim 1 / \Im(k_L)$.

\subsection{Chain with DPD damping}

Consider again a damped chain, but this time with the damping of the DPD 
form~\cite{10.1209/0295-5075/30/4/001}. In DPD, the damping depends on the 
velocity difference between nearby pairs of particles, \ahum{penalizing} motion 
that changes the pair distance (two particles moving toward each other, 
experience friction forces pointing outward, and {\it vice versa}; particles 
moving with the same velocity experience no friction). The DPD approach was 
originally designed for fluids, but is beneficial in any 
situation~\cite{10.1103/physrevlett.120.185501} where linear momentum needs to 
be conserved (recall that Langevin damping, in contrast, does {\it not} conserve 
momentum).

For the 1D chain, defining \ahum{nearby} to mean pairs of nearest neighboring 
beads, the DPD equation of motion becomes:
\begin{equation}
\label{eq:N2DPD}
\begin{split}
m \ddot{u}_i = K(u_{i-1} + u_{i+1} - 2u_i) \, &+ \\ m \gamma_D 
(\dot{u}_{i-1} &+ \dot{u}_{i+1} - 2\dot{u}_i) \,,
\end{split}
\end{equation}
where the DPD damping parameter $\gamma_D$ also has the unit of frequency. 
Mathematically, one recognizes the DPD term as the finite-difference expression 
for the second derivative (curvature) of the velocity profile. Substituting 
again the plane-wave Ansatz, and taking the limit $\omega \ll \Omega_0$, one 
obtains the wavenumber:
\begin{equation}
\label{eq:KDPD}
k_D \overset{\omega \ll \Omega_0}{\approx}
\frac{\omega}{c_0} + \frac{\gamma_D a^2 \omega^2 \, \I}{2 c_0^3 } \,.
\end{equation}
The wavenumber is again complex, meaning the wave is damped, decaying on a 
length scale $l_D \sim 1/\Im(k_D)$.

\subsection{Dissipation scaling laws: Langevin damping}

We now consider the Langevin chain under oscillatory shear driving to determine 
how the {\it dissipation}, i.e.~the energy lost on average per unit of time, 
depends on the chain length $L=aN$. To this end, bead $i=N-1$ on one chain end 
is driven, while bead $i=0$ on the other end is held fixed ($u_0=0$). Under {\it 
strain-controlled} driving, the displacement of the driving bead is prescribed:
\begin{equation}
\label{eq:sc}
u_{N-1} = A \sin (\omega t) \,, 
\end{equation}
with $\omega$ the driving frequency, and $A$ the strain amplitude. Under {\it 
force-controlled} driving, a time-dependent driving force is added to the 
driving bead:
\begin{equation}
\label{eq:fc}
F(t) = F_0 \sin (\omega t) \,,
\end{equation}
with $F_0$ the driving force amplitude. 

In the limit of low driving frequency~$\omega$, the entire chain \ahum{keeps up} 
with the driving, and a {\it quasi-static} approximation becomes reasonable. In 
this approximation, the chain is always close to its lowest energy 
configuration, which here is a straight line, implying a {\it linear} 
displacement profile:
\begin{equation}
\label{eq:lin}
u_i = A \sin(\omega t) \cdot \frac{i}{N-1} \,.
\end{equation}
The total averaged dissipation is then easily calculated. Under 
{\it strain-control}, the result is:
\begin{equation}
\label{eq:PLSC}
\begin{split}
P_L^{\rm sc} = \frac{1}{p} \int_0^p  dt \,
\sum_{i=0}^{N-1} &m \gamma_L \dot{u}_i \cdot \dot{u}_i \\
&= \frac{N (2N-1)}{12(N-1)} \cdot m \gamma_L A^2 \omega^2 \,.
\end{split}
\end{equation}
In \cref{eq:PLSC}, the summand is the product of the Langevin damping force and 
velocity of the $i$-th bead, which corresponds to power; the total dissipation 
is the sum over all beads; integrating over one driving period $p=2\pi/\omega$ 
yields the average dissipation. 

Under {\it force-control}, \cref{eq:PLSC} still applies, but with the driving 
amplitude replaced by $A \to (N-1) F_0 / K$, where $K$ is the spring constant of 
a single spring (for springs in series, the effective stiffness of the entire 
chain $K_{\rm eff} \sim K/N$~\cite{10.1088/0957-4484/22/28/285708}, and so, for 
a fixed driving force amplitude $F_0$, the strain amplitude increases with~$N$ 
because longer chains are effectively \ahum{softer}). The dissipation under 
force-control thus becomes:
\begin{equation}
\label{eq:PLFC}
P_L^{\rm fc} = 
\frac{N (2N-1)(N-1)}{12} \cdot \frac{m \gamma_L F_0^2 \omega^2}{K^2} \,.
\end{equation}

\subsection{Dissipation scaling laws: DPD damping}

Next, we provide the dissipation scaling laws for the DPD chain, again in the 
quasi-static approximation, i.e.~assuming the linear displacement profile of 
\cref{eq:lin}. In this case, under strain-controlled driving, the dissipation 
becomes:
\begin{equation}
\label{eq:PDSC}
\begin{split}
P_D^{\rm sc} &= \frac{1}{p} \int_0^p  dt \,
\gamma_D m (\dot{u}_{N-2} - \dot{u}_{N-1}) \cdot \dot{u}_{N-1} \\
&= \frac{1}{2(N-1)} \cdot m \gamma_D A^2 \omega^2 \,.
\end{split}
\end{equation}
In \cref{eq:PDSC}, one recognizes the integrand as the DPD damping force acting 
on the driven bead ($i=N-1$) multiplied by its velocity, which for a linear 
profile is the only non-zero contribution. The dissipation decreases with $N$ as 
an inverse power law, and ultimately vanishes. 

Under force-control, one again substitutes $A \to (N-1) F_0 / K$, leading to:
\begin{equation}
\label{eq:PDFC}
P_D^{\rm fc} = \frac{(N-1)}{2} \cdot 
\frac{m \gamma_D F_0^2 \omega^2}{K^2} \,,
\end{equation}
which {\it increases} linearly with~$N$. Hence, for DPD damping, the dependence 
of dissipation on chain length is crucially determined by the driving protocol, 
and can be increasing (force-control) or decreasing (strain-control).

\subsection{Range of validity}

Let us now make more precise under what conditions our {\it quasi-static} 
thickness-dissipation relations are expected to hold. In the assumed linear 
profile of \cref{eq:lin}, the beads move in phase, which is only possible if the 
wavelength $\lambda$ of the wave induced by the driving into the chain far 
exceeds the chain length~$aN$. The second condition is that the wave must be 
able to propagate along the entire chain, meaning that also the decay length~$l$ 
must exceed the chain length. Expressed mathematically, these conditions imply:
\begin{equation}
\label{eq:valid}
 \lambda = 2\pi / \Re(k) \gg aN, \quad 
 l \sim 1/\Im(k) \gg aN \,,
\end{equation}
with $k$ the wavenumber. For each situation at hand, one must check if 
\cref{eq:valid} is fulfilled, using the appropriate expression for the 
wavenumber [\cref{eq:langdamp,eq:KDPD}].

\section{1D chain: Scaling law verification}

We now present Molecular Dynamics (MD) simulations of the driven 1D chain, to 
verify the scaling laws (the simulations are performed using standard software 
tools, see Appendix for details). We choose $a \equiv 1$, $m \equiv 1$, and 
$\tau = 1/\Omega_0 \equiv 1$ as our units of length, mass, and time, 
respectively. The driving frequency is set to $\omega = 0.001 \ll \Omega_0$, 
with unit driving amplitudes $A=F_0=1$. We do not include any random thermal 
forces at this point, so the presented results correspond strictly to zero 
temperature. In the simulations, the dissipation is taken to be the rate of work 
done driving the top bead $i=N-1$, which takes the general form:
\begin{equation}
\label{eq:psim}
\PMD = \expval{ \tilde{F}(t) \, \tilde{V}(t) }_t \,,
\end{equation}
where $\avg{\cdot}_t$ denotes a time-average. 

In the simulations, $\PMD$ is obtained from the trajectory, i.e.~the time series 
of the bead displacements $u_i^{\rm MD} (t)$ and velocities $\dot{u}_i^{\rm MD} 
(t)$, where $t$ is time. Under strain-control, the velocity of the driving bead 
is prescribed by \cref{eq:sc}, implying $\tilde{V}(t) = \dot{u}_{N-1}$, the 
force being $\tilde{F} (t) = -F_{N-1}^{\rm MD}$, i.e.~the sum of elastic and 
damping forces exerted on the driving bead as obtained from the trajectory (the 
minus sign is convention, such that in a dissipative system, $\PMD>0$). Under 
force-control, the driving force is prescribed by \cref{eq:fc}, 
$\tilde{F}(t)=F(t)$, with now the velocity of the driving bead $\tilde{V}(t) = 
\dot{u}_{N-1}^{\rm MD}$ taken from the trajectory. When collecting the time 
average, the total simulation duration should span an integer multiple of 
driving periods, with the first few periods discarded to allow for any initial 
transients to vanish.

\begin{figure}
%% m2gk/1D_chain/fig_langevin.py
\centering
\includegraphics[width=0.95\columnwidth]{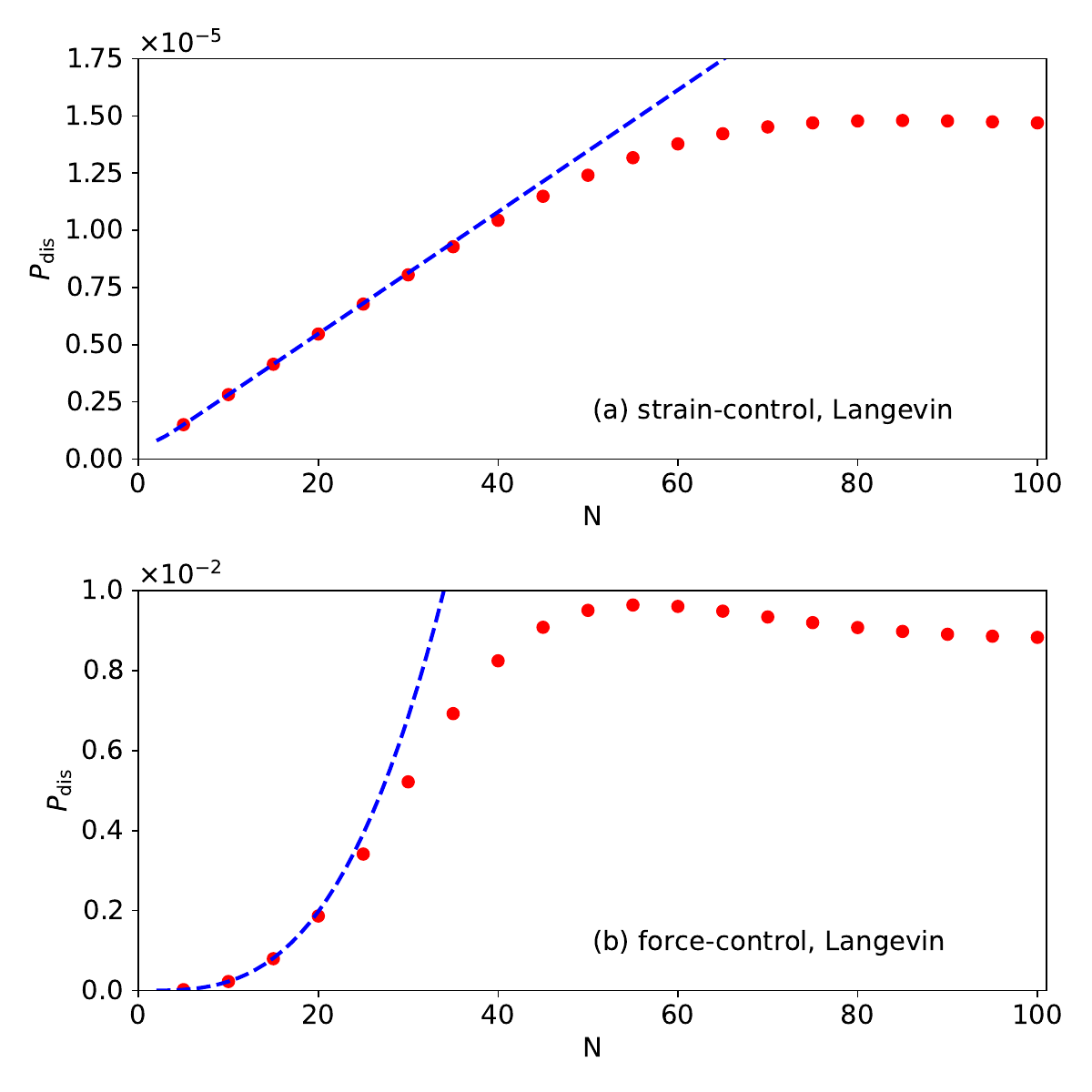}
\caption{\label{fig:langevin} Dissipation $\PMD$ versus chain length $N$, for 
the 1D chain with Langevin damping, under (a) strain- and (b) force-controlled 
driving. The dashed lines show the appropriate asymptotic scaling law, given by 
\cref{eq:PLSC,eq:PLFC}, respectively.}
\end{figure}

\begin{figure}
%% m2gk/1D_chain/fig_dpd.py
\centering
\includegraphics[width=0.95\columnwidth]{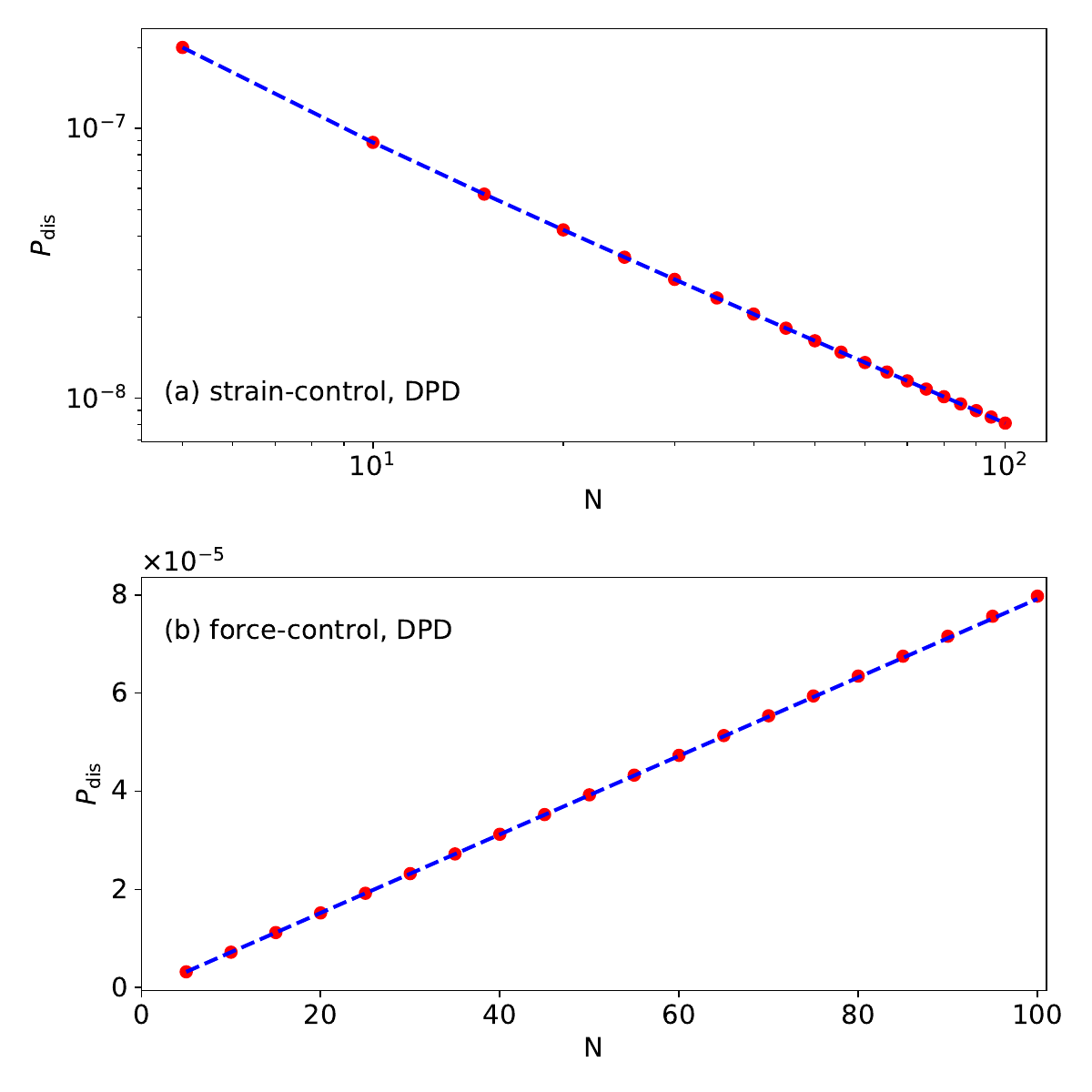}
\caption{\label{fig:dpd} Analogue of \cref{fig:langevin}, using DPD damping, 
with scaling laws given by \cref{eq:PDSC,eq:PDFC}. Note that panel (a) 
uses double-logarithmic scales.}
\end{figure}

We first consider the Langevin chain using damping parameter $\gamma_L=1.6 \gg 
\omega$. \cref{fig:langevin}(a) shows $\PMD$ for strain-controlled driving 
versus the chain length~$N$, as obtained using MD (dots). Applying now the 
criteria of \cref{eq:valid} with \cref{eq:langdamp}, our dissipation scaling 
laws should be valid provided $N < \Nstar \sim c_0 / \sqrt{\gamma_L\omega} \sim 
25$. The dashed line shows the scaling law, \cref{eq:PLSC}, which, in the regime 
where $N<\Nstar$, correctly captures the data (we emphasize that the dashed line 
is not a fit, since all the parameters in \cref{eq:PLSC} are known). For 
$N>\Nstar$, the wave induced by the driving is so strongly damped, it no longer 
is able to reach the other chain end. In this case, adding more beads does not 
affect the dissipation, and so $\PMD$ saturates. \cref{fig:langevin}(b) shows 
the result for force-controlled driving, the dashed line here corresponding to 
\cref{eq:PLFC}, which, in the regime $N<\Nstar$, also correctly captures the 
data.

In \cref{fig:dpd}, we show the analogous results for the DPD chain, using 
$\gamma_D=1.6$. Following \cref{eq:KDPD}, the wave length $\lambda = 2\pi / 
\Re(k_D) \sim 6300$ and decay length $l_D \sim 1/\Im(k_D) > 10^6$, which both 
far exceed our considered chain lengths. In line with the criteria of 
\cref{eq:valid}, the DPD scaling laws thus capture the entire data range. Note 
the pronounced qualitative difference: decreasing dissipation with $N$ for 
strain-control, increasing for force-control.

\section{Dissipation in a 3D cubic crystal}
\label{sec:3d}

\begin{comment}
SUMMARY OF ELASTIC CONSTANTS FOR 3D CUBIC SYSTEM:
Bulk Modulus = 3.33333333324243 LJ-units
Shear Modulus 1 = 1.99999999997687 LJ-units
Shear Modulus 2 = 1.99999999992817 LJ-units
Poisson Ratio = 0.250000000001801
Young = 3*BULK*(1-2*Poisson) = 2*SHEAR*(1+Poisson) = 5
See also: http://dx.doi.org/10.1140/epjb/e2015-60329-5
\end{comment}

We now consider a 3D crystal, using essentially the particle model of 
\olcite{10.1103/PhysRevB.104.174309}. The crystal structure is taken to be {\it 
simple-cubic}, the lattice constant $a \equiv 1$ being our unit of length. We 
use simulation cells of size $N_x=N_y=20$ in the two lateral directions; the 
number of vertical layers is denoted~$N_z$. Periodic boundary conditions are 
applied in the lateral directions, but not in the vertical one. Between nearest- 
and next-nearest neighboring atoms harmonic bonds are placed, both with spring 
constant~$K$, the rest-lengths of the springs being $1$ and $\sqrt{2}$, 
respectively, such that the perfect crystal configuration has zero energy. We 
again take the single particle mass to be unity, $m \equiv 1$, and as unit of 
time $\tau = 1 / \Omega_0 = \sqrt{m/K} \equiv 1$. In these units, the 
(zero-temperature) shear modulus of the crystal $G=2$, Young modulus $E=5$, and 
Poisson ratio $\nu=0.25$ (which were obtained numerically by deforming the 
crystal cell and energy minimization).

The bottom layer of the crystal is kept fixed, while the top layer is 
harmonically driven in the $x$-direction with frequency $\omega=2\pi \cdot 
0.001$, corresponding to a global oscillatory shear deformation. We consider both 
{\it strain}- and {\it force}-controlled scenarios, the respective driving 
amplitudes being $A=F_0=0.05$. For strain-controlled driving, the 
$x$-displacement of each atom in the top layer is prescribed, $u_x = A \sin 
(\omega t)$, with $u_x$ measured from the perfect lattice position (the 
remaining $y,z$ components are allowed to move freely under the influence of 
elastic and damping forces). For force-controlled driving, atoms in the top 
layer have an additional force component in the $x$-direction, $F_x = F_0 \sin 
(\omega t)$, added to them. The dynamics of the system is obtained using MD (see 
Appendix for details). The dissipation is again measured using \cref{eq:psim}, 
but divided by $N_xN_y$, so that the reported dissipation $\PMD$ is here to be 
understood as the dissipation {\it per surface atom}.

\begin{figure}
%% m2gk/full_shear/fig_langevin.py
\centering
\includegraphics[width=0.95\columnwidth]{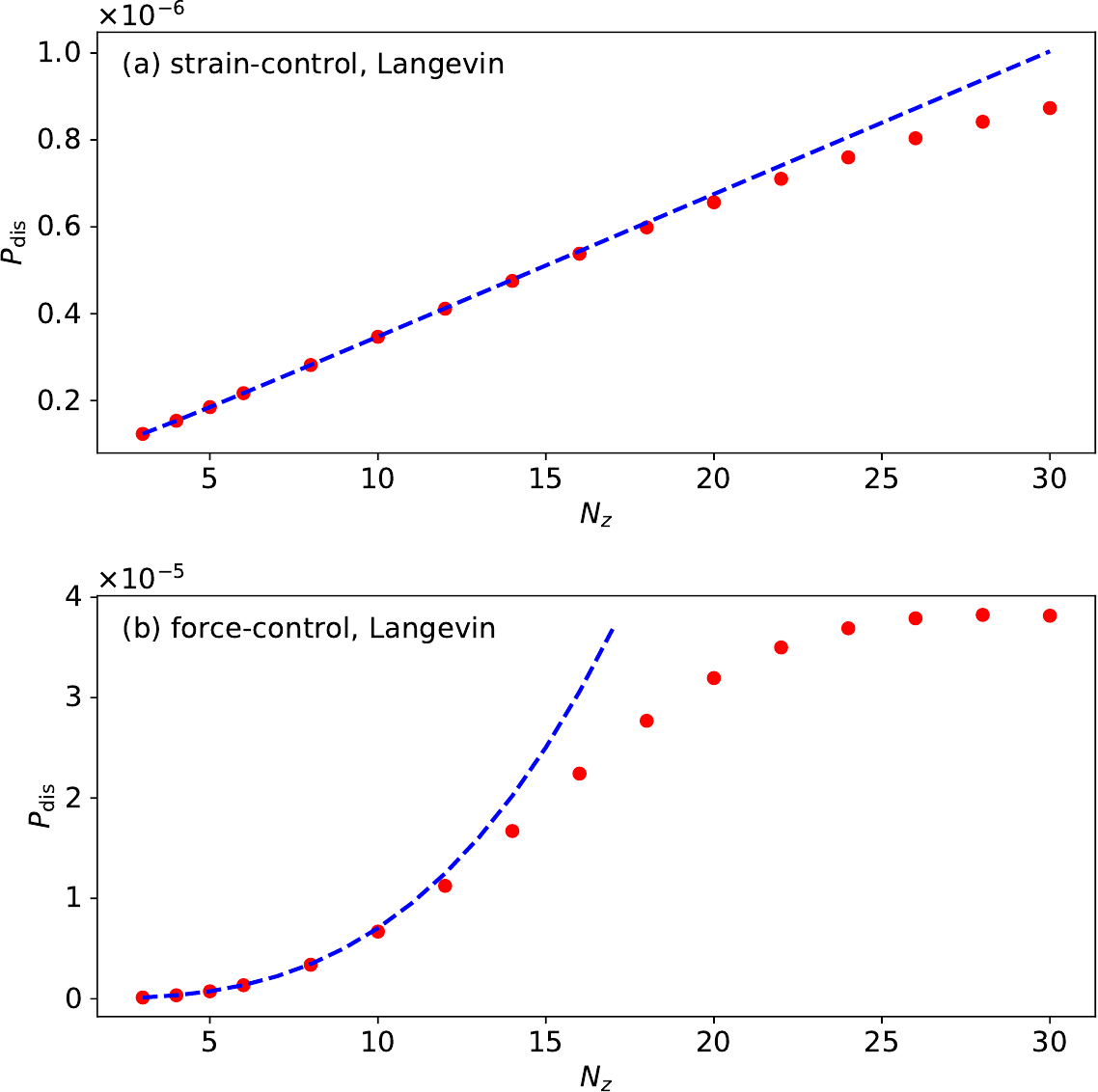}

\caption{\label{fig:l3d} Dissipation $\PMD$ (per surface atom) versus the number 
of vertical layers~$N_z$, for a 3D cubic crystal with Langevin damping, under 
(a) strain- and (b) force-controlled driving. Dots show simulation results, 
dashed lines the appropriate asymptotic scaling law, given by 
\cref{eq:PLSC,eq:PLFC}, respectively.}

\end{figure}

\subsection{Zero temperature}

We first consider the crystal with Langevin damping at zero temperature. In 3D, 
the Langevin damping force acting on particle $i$ is given by $\vec{F}_{L,i} = 
-m \gamma_L \vec{v}_i$, with $\vec{v}_i$ the velocity {\it vector} of 
particle~$i$. With the exception of the frozen bottom layer, the damping force 
is applied to every particle in the system, including those being driven, using 
$\gamma_L = 2 \gg \omega$. Following \cref{eq:langdamp,eq:valid}, our 
dissipation scaling laws should hold provided $N_z < \Nstar_z \sim 13$. In 
\cref{fig:l3d}, we show the dissipation as a function of the number of vertical 
layers $N_z$ for (a) strain- and (b) force-controlled driving. The dashed lines 
show \cref{eq:PLSC,eq:PLFC} with, in the latter, $K$ replaced by the shear 
modulus~$G$. Again, these lines are not fits, since all the required quantities 
are known. For small $N_z$, the agreement is excellent; deviations appear when 
$N_z \sim \Nstar_z$, as expected. Our observations imply that each crystal layer 
effectively moves as a single entity, i.e.~can be treated as one massive bead. 
Since the driving amplitude here is small, only shear motion is induced, the 
coupling to longitudinal motion being negligible, which explains why the crystal 
behaves effectively as a 1D chain.

\begin{figure}
%% m2gk/full_shear/fig_dpd.py
\centering
\includegraphics[width=0.95\columnwidth]{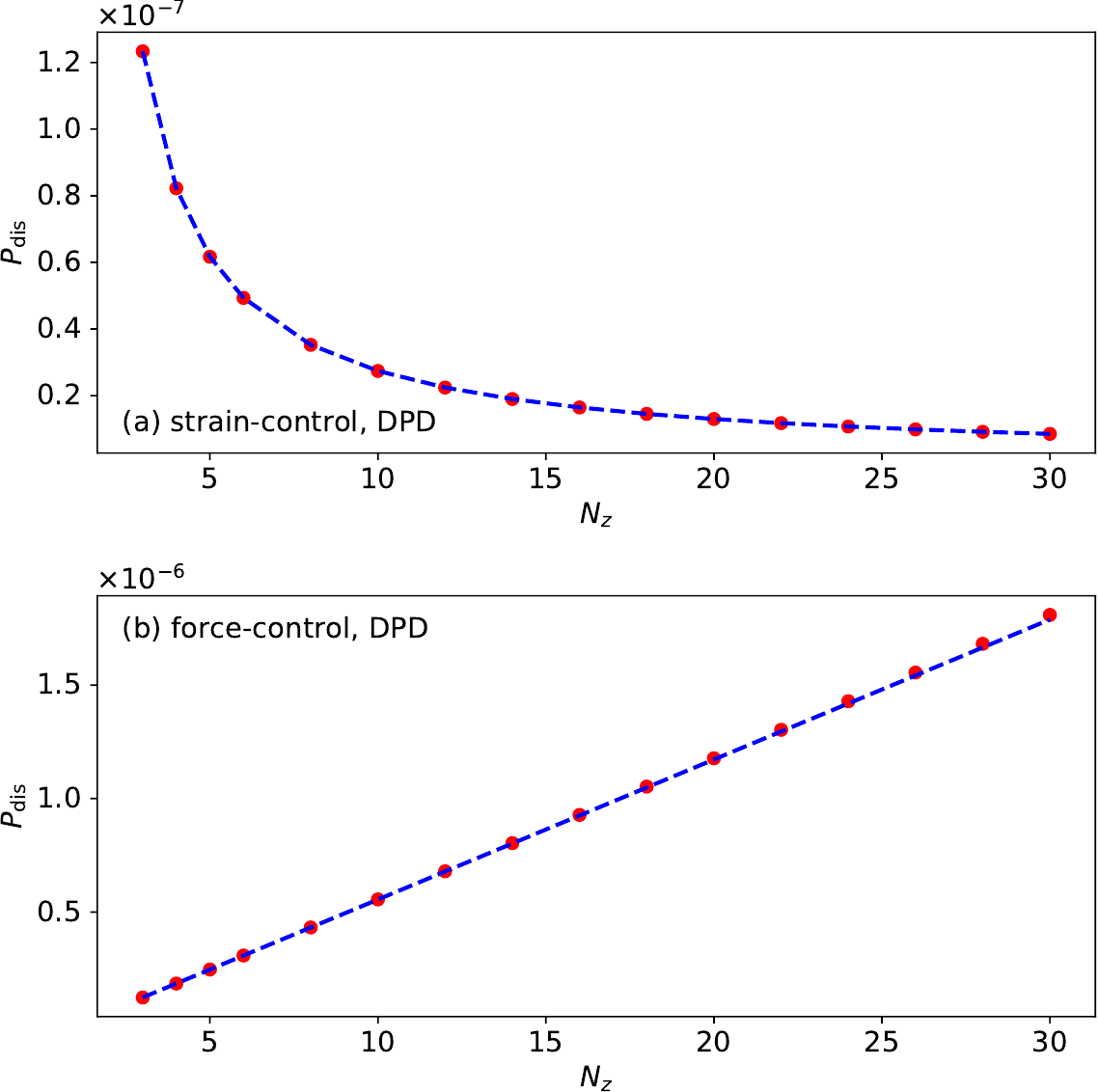}
\caption{\label{fig:d3d} Analogue of \cref{fig:l3d}, using DPD damping, with 
scaling laws given by \cref{eq:PDSC,eq:PDFC}.}
\end{figure}

Next, we consider the crystal with DPD damping at zero temperature. The DPD 
damping force is computed for every bond in the system (nearest- and 
next-nearest neighbors). The corresponding forces are applied to all particles, 
including driven ones, but excluding the bottom layer. In 3D, the DPD damping 
force acting on particle $i$ due to a bonded neighbor $j$ is given by 
$\vec{F}_{D,i} = m \gamma_D \, \hat{r}_{ij} \cdot ( \vec{v}_j - \vec{v}_i ) \, 
\hat{r}_{ij}$~\cite{10.1209/0295-5075/30/4/001}, with $\hat{r}_{ij}$ the unit 
vector pointing from particle $i \to j$ (due to Newton's third law, particle $j$ 
feels the same force acting in the opposite direction, $\vec{F}_{D,j} = - 
\vec{F}_{D,i}$). The total damping force on any given particle is obtained by 
summing over its bonded neighbors (some of which may involve frozen particles of 
the bottom layer). In \cref{fig:d3d}, we show the dissipation as a function of 
the number of vertical layers $N_z$ for (a) strain- and (b) force-controlled 
driving, using damping parameter $\gamma_D=5$. The dashed lines show 
\cref{eq:PDSC,eq:PDFC}, with $K$ again replaced by the shear modulus~$G$. 
Following \cref{eq:KDPD}, one sees that the wavelength and the decay length both 
far exceed the simulated thicknesses ($\lambda, l_D \gg a N_z$), and so, in line 
with \cref{eq:valid}, the scaling laws capture the full data range. Note again 
that the dissipation decreases with $N_z$ under strain-control, and that it 
increases under force-control.

\begin{figure}
%% m2gk/full_shear/fig_ft.py
\centering
\includegraphics[width=0.95\columnwidth]{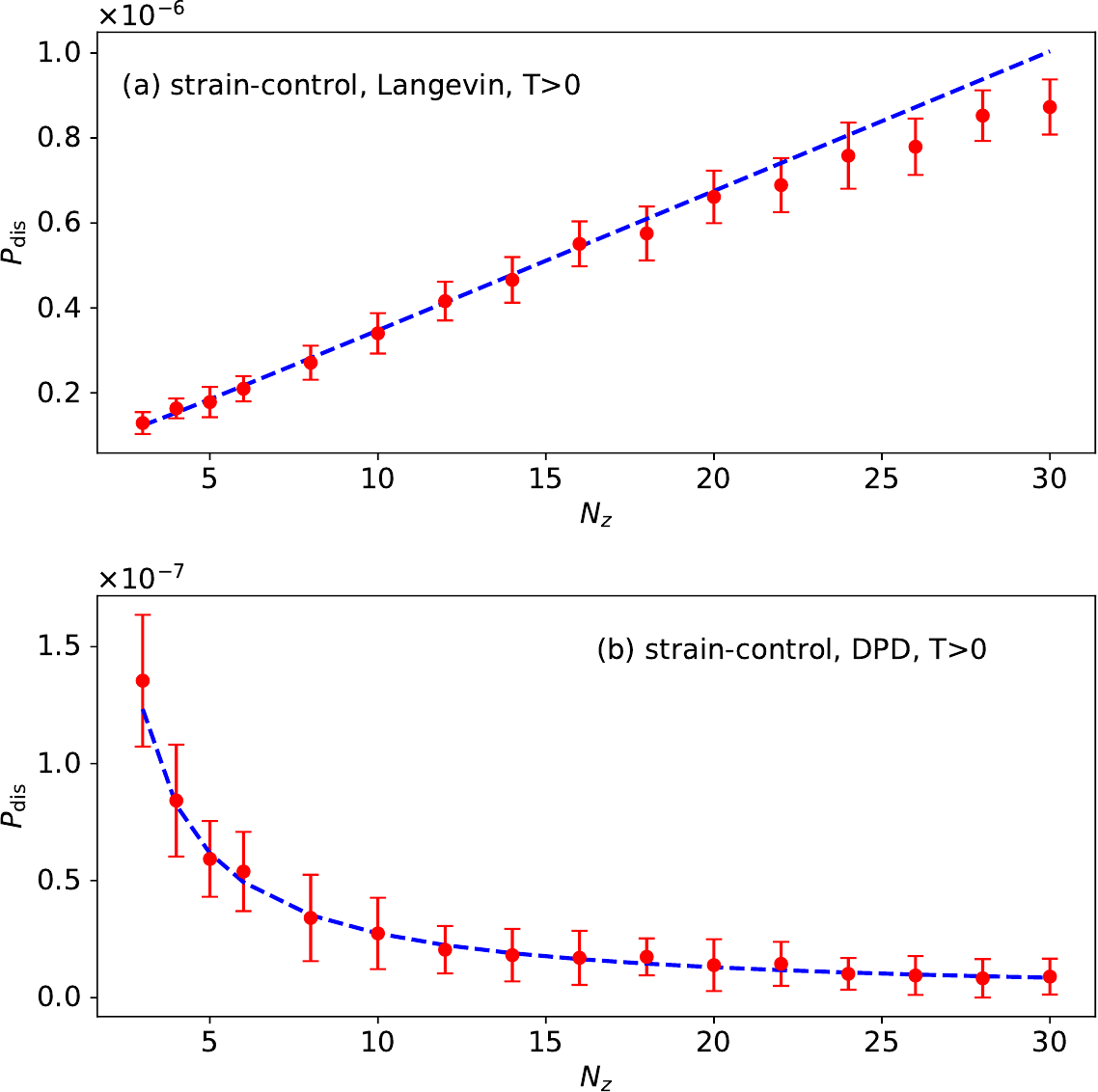}
\caption{\label{fig:ft} Numerically obtained dissipation (dots) under strain 
control for a 3D cubic crystal at {\it finite temperature} with (a) Langevin and 
(b) DPD damping versus the number of vertical layers $N_z$ (dots show the 
dissipation averaged over 50 driving periods; error bars indicate the 
corresponding standard deviation). The dashed lines show \cref{eq:PLSC,eq:PDSC}, 
respectively.}
\end{figure}

\subsection{Finite temperature}

The results presented so far apply to zero temperature, but we expect agreement 
at finite temperature also, provided one remains in the solid phase. For a 
typical metal at room temperature, the ratio of thermal to elastic energy $k_BT 
/ mc_0^2 \sim 10^{-3}$ (Boltzmann constant~$k_B$), with $c_0$ the transversal 
speed of sound, i.e.~elasticity still dominates. To verify, we have repeated two 
of our MD~runs with random thermal forces included, using $k_BT = 0.005$ in our 
energy units, while keeping all other parameters the same. For the Langevin 
system, random forces were implemented following 
\olcite{10.1142/s0129183191001037}; for DPD, the momentum-conserving noise term 
of \olcite{10.1209/0295-5075/30/4/001} was used. Results are shown in 
\cref{fig:ft}, using strain-controlled driving, for the Langevin (a) and DPD 
crystal (b). Provided the simulation spans sufficiently many driving periods, 
the average dissipation (dots) remains well described by the \ahum{$T=0$} 
scaling laws. However, there is a sizable thermal fluctuation (error bars) 
meaning that, for a single cycle, there can be considerable deviations from 
these laws (by increasing the {\it lateral} system size, $N_xN_y \to \infty$, we 
expect these fluctuations to vanish though).

\section{Summary and Conclusions}

We studied dissipation in solids subjected to global oscillatory shear as a 
function of sample thickness. We considered the effect of the damping mechanism 
(Langevin vs.~Dissipative Particle Dynamics) as well as that of the driving 
protocol (strain- vs.~force-controlled). Depending on these, dissipation can 
either increase or decrease with sample thickness. These findings can be 
understood physically using the 1D harmonic chain as model to describe the 
solid. Possible {\it experimental} verification to determine which damping 
mechanism is the relevant one, Langevin or DPD, could be performed by measuring 
the dissipation for various sample thicknesses under global oscillatory shear 
for a known driving protocol.

Regarding the use of DPD to describe the electron-phonon coupling in 
metals~\cite{10.1103/physrevlett.120.185501}, we still provide the typical value 
of the damping parameter $\gamma_D$ that should be used. Based on the number 
provided in \olcite{10.1103/physrevlett.120.185501}, the DPD damping parameter 
$\gamma_D \sim \alpha^2/m \sim 1.6 \, \rm THz$ (this uses $\alpha^2 = 0.01 \, 
\rm eV \, ps / \AA^2$ for Nickel provided in 
\olcite{10.1103/physrevlett.120.185501}, with $m$ the Nickel atomic mass). It is 
interesting to see what attenuation length this number implies. Following 
\cref{eq:KDPD} and assuming ultrasound driving ($\omega \sim 10 \, \rm MHz$), 
the above value of $\gamma_D$, together with $c_0 \sim 3000 \, \rm m/s$ and 
lattice constant $a = 3.5 \, \rm \AA$, yield a decay (attenuation) length $l_D 
\sim 1 \, \rm m$. This would be the attenuation in a perfect crystalline sample, 
and with electron-phonon coupling being the only dissipation channel. The 
actually measured attenuation in real metals in the ultrasound regime is 
typically $\sim 100 \, \rm dB/m$~\cite{10.3390/app10072230}, corresponding to an 
attenuation length of centimeters, i.e.~significantly smaller. This shows that, 
in real metals, other mechanisms (besides electron-phonon coupling) also are at 
play, enhancing dissipation~\cite{10.1063/1.1722284, 10.1103/physrev.173.856}.

We still illustrate a delicate point that arises when damping schemes are used 
to model real materials, say, for the purpose of MD simulation. As stated 
before, any such scheme requires that the damping parameter be specified. For 
Langevin and DPD, these parameters are $\gamma_L$ and $\gamma_D$, respectively. 
A sensible approach might seem to fit these parameters to the attenuation 
length. For any given driving frequency and for each damping scheme (Langevin or 
DPD), one can always select the damping parameter to match a desired attenuation 
length, using \cref{eq:langdamp,eq:KDPD}. However, the corresponding dissipation 
between the damping schemes will be very different, see for example 
\cref{fig:ft}, even if the attenuation is the same! In other words, mere 
agreement with the attenuation length is no guarantee that also the dissipation 
will be captured correctly. Instead, additional microscopic information is 
needed to select the appropriate damping model.

\acknowledgments

This work was supported by the Deutsche Forschungsgemeinschaft (DFG, German
Research Foundation) -- 217133147/SFB 1073, project A01.

\appendix

\section{MD Simulation details}

The 1D chain simulations were performed by integrating the equations of motion 
using the Runge-Kutta-Fehlberg method of the GNU Scientific Library~\cite{gsl} 
with integration timestep $\Delta t =0.005$. The MD simulations of the 3D 
crystal were performed with LAMMPS~\cite{10.1016/j.cpc.2021.108171} using 
integration time step $\Delta t = 0.001$. To implement DPD bonds, a custom bond 
style was coded; all other aspects of the simulations can be modeled with 
LAMMPS using its standard features.

\bibliography{STUFF,AUTODOI}

\end{document}